\documentstyle[12pt]{article}

\renewcommand{\theequation}{\thesection.\arabic{equation}}
\begin{document}
\newcommand{\pl}[1]{Phys.\ Lett.\ {\bf #1}\ }
\newcommand{\npb}[1]{Nucl.\ Phys.\ {\bf B#1}\ }
\newcommand{\prd}[1]{Phys.\ Rev.\ {\bf D#1}\ }
\newcommand{\prl}[1]{Phys.\ Rev.\ Lett.\ {\bf #1}\ }

\newcommand{\PSbox}[3]{\mbox{\rule{0in}{#3}\includegraphics{#1}\hspace{#2}}}

\newcommand{\drawsquare}[2]{\hbox{%
\rule{#2pt}{#1pt}\hskip-#2pt
\rule{#1pt}{#2pt}\hskip-#1pt
\rule[#1pt]{#1pt}{#2pt}}\rule[#1pt]{#2pt}{#2pt}\hskip-#2pt
\rule{#2pt}{#1pt}}

\newcommand{\Yfund}{\raisebox{-.5pt}{\drawsquare{6.5}{0.4}}}
\newcommand{\Ysymm}{\raisebox{-.5pt}{\drawsquare{6.5}{0.4}}\hskip-0.4pt%
        \raisebox{-.5pt}{\drawsquare{6.5}{0.4}}}
\newcommand{\Yasymm}{\raisebox{-3.5pt}{\drawsquare{6.5}{0.4}}\hskip-6.9pt%
        \raisebox{3pt}{\drawsquare{6.5}{0.4}}}
\newcommand{\Ythree}{\raisebox{-3.5pt}{\drawsquare{6.5}{0.4}}\hskip-6.9pt%
        \raisebox{3pt}{\drawsquare{6.5}{0.4}}\hskip-6.9pt
        \raisebox{9.5pt}{\drawsquare{6.5}{0.4}}}

\newcommand{\beq}{\begin{equation}}
\newcommand{\eeq}{\end{equation}}
\newcommand{\beqa}{\begin{eqnarray}}
\newcommand{\eeqa}{\end{eqnarray}}

\newcommand{\jref}[4]{{\it #1} {\bf #2}, #3 (#4)}
\newcommand{\NPB}[3]{\jref{Nucl.\ Phys.}{B#1}{#2}{#3}}
\newcommand{\PLB}[3]{\jref{Phys.\ Lett.}{#1B}{#2}{#3}}
\newcommand{\PR}[3]{\jref{Phys.\ Rep.}{#1}{#2}{#3}}
\newcommand{\PRD}[3]{\jref{Phys.\ Rev.}{D#1}{#2}{#3}}
\newcommand{\PRL}[3]{\jref{Phys.\ Rev.\ Lett.}{#1}{#2}{#3}}
\newcommand{\PRV}[3]{\jref{Phys.\ Rev.}{#1}{#2}{#3}}

\renewcommand{\thefigure}{\arabic{figure}}
\setcounter{figure}{0}
\begin{titlepage}
\begin{center}
{\hbox to\hsize{hep-th/9906057   \hfill  LBNL-43423}}
{\hbox to\hsize{               \hfill  UCB-PTH-99/27}}
{\hbox to\hsize{               \hfill  UCSD-PTH-99/07}}
\bigskip

\bigskip

{\Large \bf  $\beta$ Functions of Orbifold Theories and the
Hierarchy Problem}

\bigskip

\bigskip

{\bf Csaba Cs\'aki$^{a,b,}$\footnote{Research fellow, Miller Institute for
Basic Research in Science.}, Witold Skiba$^{c}$, and
John Terning$^{a,b}$}\\

\smallskip

{\small \it $^a$Department of Physics, University of California, Berkeley,
CA 94720}

\smallskip

{\small \it $^b$Theoretical Physics Group,
Lawrence Berkeley National Laboratory\\
     University of California, Berkeley, CA 94720}

\smallskip

{\small \it $^c$Department of Physics, University of California at San Diego,\\
 La Jolla,  CA 92093}

\bigskip

{\tt  ccsaki@lbl.gov, skiba@einstein.ucsd.edu, terning@alvin.lbl.gov}

\bigskip

\vspace*{1cm}
{\bf Abstract}\\
\end{center}
We examine a class of gauge theories obtained by projecting out
certain fields from an ${\cal N}=4$ supersymmetric
$SU(N)$ gauge theory. These theories are non-supersymmetric and
in the large $N$ limit are known to be
conformal. Recently
it was proposed that the hierarchy problem could be solved by
embedding the standard model in a theory of this kind with finite $N$.
In order to check this claim one must find the conformal points of the
theory. To do this we calculate the
one-loop $\beta$ functions for the Yukawa and quartic scalar couplings.
We find that with the $\beta$ functions set to zero the one-loop quadratic
divergences are not canceled at sub-leading order in $N$; thus the
hierarchy between the weak scale and the Planck scale
is not stabilized unless $N$ is of the order $10^{28}$ or larger.
We also find that at sub-leading orders in $N$ renormalization
induces new interactions, which were not present in the original Lagrangian.

\bigskip

\bigskip

\end{titlepage}

\section{Introduction}
\setcounter{equation}{0}
\setcounter{footnote}{0}

The study of conformal symmetry has a long history in particle physics.
Recently it has attracted renewed interest due to the work of
Maldacena~\cite{Maldacena}
on the correspondence between string theory on anti-de Sitter backgrounds
and four dimensional conformal field theories, and further work on
the orbifold projections of these theories~[2-12]. An interesting
result of this work~[2-6] is that
non-supersymmetric gauge theories obtained by orbifolding an
${\cal N}=4$ SUSY $SU(N)$ gauge theory are conformal in the large $N$
limit. Additional non-supersymmetric conformal theories can be obtained
from a similar construction in type 0 string theories~\cite{Type0,Nekrasov}.
Although conformal theories
are seemingly quite esoteric, the idea of using
static or slowly running couplings to generate a large hierarchy of scales has
cropped up many times in particle phenomenology.
Attempts to use approximate conformal symmetry in phenomenology have included
such diverse topics as: electroweak symmetry breaking (walking technicolor)
\cite{walking,Yamawaki,Postmodern},  the
hunt for light composite scalars \cite{Yamawaki,dilatons,lightscalar}
 (including the search for  a Goldstone boson of spontaneously
broken scale invariance\footnote{The relation between scale invariance
and conformal invariance is discussed in Ref.~\protect\cite{Polchinski}.}
\cite{dilatons}),
dynamical supersymmetry breaking \cite{NelsonStrassler}, and  the
cosmological constant problem \cite{Kachru}.
Most recently Frampton and Vafa
\cite{VafaFramp,Frampton} have
conjectured that orbifold theories are conformal at finite $N$, and
further proposed that embedding the standard model in an orbifold theory can
solve
the naturalness problem
of the electroweak scale (stabilizing the large hierarchy of scales without
fine-tuning).
This sudden appearance of such a simple solution to a long standing problem  is
quite surprising, so it seems worthwhile to discuss the underlying ideas of
this
scenario in some detail.

It has been previously noted \cite{Bardeen} that conformal symmetry can
remove the quadratic divergences that are responsible for destabilizing the
hierarchy between the
weak scale and a more fundamental scale like the Planck scale.  In a
conformal theory
we must insist on regulators (like
dimensional regularization) that respect conformal invariance
or include counterterms that maintain the symmetry. With such a
regularization quadratic divergences are impossible (since there is no cutoff
scale on which they could depend).  Such a resolution of the naturalness
problem is of course only valid if the theory
is exactly conformal (i.e. physics is the same at any length scale).  In the
real world we know that
physics is not conformal below the weak scale, and we expect that the
fundamental theory of
gravity will not be scale invariant since gravity has an intrinsic scale
associated with it.  Thus the best we can hope for phenomenologically is a
theory that is approximately scale invariant in some energy range.  That is
we can only have an effective conformal theory that is valid above
some infrared cutoff (which must be above the weak scale) and below some
ultraviolet (UV) cutoff $M$ (which must be at or below the Planck scale).  From
the perspective of the fundamental theory there is some non-conformal physics
above (or near) the scale $M$ (e.g. heavy particles or massive string modes)
which we can integrate out of the theory.  Studying the sensitivity of the
effective theory to the cutoff $M$ is equivalent to studying
sensitivity of the low-energy
physics to the details of the very high-energy physics. If we believe that
there is a new fundamental scale of physics beyond the weak scale then in a
``natural theory" we would like to
see that the weak scale is not quadratically sensitive to changes in the
high scale.  The two known solutions to the naturalness problem are to either
lower the UV cutoff of the effective
theory to the weak scale (e.g. technicolor, large extra dimensions) or to
arrange cancelations of the quadratic divergences order-by-order in
perturbation theory (e.g. supersymmetry).  One might expect that an effective
conformal theory would fall into the latter category, however the vanishing
of $\beta$ functions does not imply the cancellation of quadratic divergences,
 they  are independent \cite{Bardeen}.  To see that they are independent
one need only consider
supersymmetric theories where quadratic divergences cancel independently of
the values of $\beta$ functions.

In this paper we consider a class of  ${\cal N}=4$ orbifold 
theories~\cite{VafaFramp,Frampton} at one loop. 
We explicitly calculate the $\beta$  functions,
solve for the couplings by imposing that the $\beta$ functions vanish,
and calculate the quadratic divergences.  We find that the quadratic
divergences do not cancel for finite $N$. We also discuss new interactions
that are induced by renormalization group (RG) running, and remark on some open
questions.

\section{The Orbifold Theories}
\setcounter{equation}{0}
\setcounter{footnote}{0}

In this section we review the construction of ${\cal N}=4$ orbifold
theories, and present the matter content and Lagrangian for the
particular models that we will be considering in this paper.

One starts with an ${\cal N}=4$ supersymmetric $SU(N)$ gauge theory. The field
content of this theory is (all fields are in the adjoint representation):
gauge bosons $A_{\mu}$, which are singlets of the $SU(4)_R$ global
symmetry, four copies of (two-component) Weyl fermions $\Psi^i,i=1,2,3,4$,
which transform as the fundamental {\bf 4} under the $SU(4)_R$,
and six copies of (real) scalars $\Phi^{ij}$ which transform as the
antisymmetric tensor {\bf 6} of $SU(4)_R$. In the procedure of
orbifolding (discussed in detail in Refs.~[2-10]) one chooses a discrete
subgroup $\Gamma$ of the $SU(4)_R$ symmetry of order $|\Gamma |$, and also
embeds this subgroup into the gauge group
(chosen here to be $SU(N|\Gamma |)$)
as $N$ copies of its regular representation (for a very clear explanation of
this embedding see Ref.~\cite{Schmaltz}). Orbifolding then means projecting out
all fields from the theory which are not invariant under the action
of the discrete group $\Gamma$. If $\Gamma$ is a generic subgroup of $SU(4)_R$,
then one obtains a non-supersymmetric theory. If $\Gamma$ is embedded in an
$SU(3)$ subgroup of $SU(4)_R$
then one obtains an  ${\cal N}=1$ supersymmetric theory,
while if  $\Gamma$ is embedded in an
$SU(2)$ subgroup of $SU(4)_R$
one obtains an  ${\cal N}=2$ supersymmetric theory.
For a compilation of results on discrete
subgroups of $SU(3)$ and $SU(4)$ see Refs.~\cite{Hanany1} and \cite{Hanany2}.
We are interested only in the
non-supersymmetric theories, in which case $\Gamma $ must be a subgroup
of $SU(4)$. In order to simplify the analysis of the $\beta$ functions,
we restrict our attention in this paper to the case when $\Gamma$ is Abelian,
$\Gamma =Z_k$. In this case we start with an $SU(Nk)$ gauge group,
and after orbifolding we obtain an $SU(N)^k$ theory.

Let us denote the $k$-th root of unity $e^{\frac{2\pi i}{k}}$
by $\omega$. An embedding of
$Z_k$ into $SU(4)_R$ is specified by the transformation properties
of the fundamental representation: ${\bf 4}\to {\rm diag}\,
(\omega^{k_1},\omega^{k_2},
\omega^{k_3},\omega^{k_4})\, {\bf 4}$. This embedding is an $SU(4)$
subgroup if
$k_1+k_2+k_3+k_4=0 \ {\rm mod} \ k$ (in order to insure that the
determinant is one), moreover $k_1,k_2,k_3,k_4\neq 0 \ {\rm mod}\  k$
so that we obtain a non-supersymmetric theory.
In order to simplify our calculations, we will assume in this paper
that no two $k_i$'s are equal, and also that $k_i+k_j\neq 0\ {\rm mod}\ k$.
With the assumption that $k_i+k_j\neq 0 \ {\rm mod} \ k$
one can avoid the presence of adjoint scalars, and
thus all fermions and scalars will be in bifundamental representations.
The assumption that no two $k_i$'s are equal implies that there is
only a single field with given gauge quantum numbers.
This is probably the simplest and most symmetric orbifold theory that one
can consider. However, we believe that the conclusions we draw from
these particular orbifolds could be generalized to more complicated embeddings.

With this choice of embedding of the discrete group
we get the following field content for our orbifold theory: \\
\indent
- gauge bosons $A_{\mu}$ for every gauge group $SU(N)$, \\
\indent
- (two-component) fermions $\Psi (m,\overline{m+k_i})\equiv
                                   \Psi^{\alpha_m}_{\beta_{m+i}}$
which transform as fundamentals under the $m$-th $SU(N)$ factor
in the $SU(N)^k$ product and as antifundamental under $m+k_i$
($m$ is arbitrary, $i=1,2,3,4$, and $m+i$ is a short hand for $m+k_i$), \\
\indent
- complex scalars $\Phi (m,\overline{m+l_i})\equiv\Phi^{\alpha_m}_{
\beta_{m+i}}$ which transform as
fundamentals under the $m$-th group and as antifundamentals under
$m+l_i$ ($m$ is arbitrary, $l_i=k_i+k_4$ and $i=1,2,3$). Note that
for the scalars a different shorthand is employed, $m+i$ represents
$m+k_i+k_4$.

The Lagrangian of orbifold theories is obtained from the original
Lagrangian by retaining only terms containing fields invariant under
the discrete symmetry. We give the ${\cal N}=4$ Lagrangian in
the Appendix. The Yukawa couplings in the orbifold theory are given by
\begin{equation}
{\cal L}_{Yukawa}= -Y\sum_{m,i<j} \left( \Psi^{\alpha_m}_{\beta_{m+i}}
\Psi_{\gamma_p}^{\beta_{m+i}}
{\Phi^{\dagger}}^{\gamma_p}_{\alpha_m} +h.c. \right),
\end{equation}
where in the above sum $m+i$ is again a shorthand for $m+k_i$, and
$p=m+k_i+k_j$.
Note that, unlike in the supersymmetric theory, there is
no factor of $\sqrt{2}$ appearing in this coupling.
The quartic scalar couplings are given by

\beqa
\label{Lagrangian}
{\cal L}_{\rm quart}&=& -\frac{1}{2} \sum_{m,j<i} \left[
\lambda_1 \, \phi^{\alpha_m}_{\beta_{m+i}}  \phi^{\gamma_{m+i}}_{\alpha_m}
\phi^{\delta_m}_{\gamma_{m+i}} \phi^{\beta_{m+i}}_{\delta_m}
-
\lambda_3 \,  \phi^{\alpha_m}_{\beta_{m-i}} \phi^{\gamma_{m+i}}_{\alpha_m}
\phi^{\delta_m}_{\gamma_{m+i}} \phi^{\beta_{m-i}}_{\delta_m} \right.
\nonumber \\
&&+
\lambda_4 \, \left( \phi^{\alpha_m}_{\beta_{m-i}}
\phi^{\gamma_{m+j}}_{\alpha_m}
\phi^{\delta_m}_{\gamma_{m+j}} \phi^{\beta_{m-i}}_{\delta_m}
+ \phi^{\alpha_m}_{\beta_{m+i}} \phi^{\gamma_{m-j}}_{\alpha_m}
\phi^{\delta_m}_{\gamma_{m-j}} \phi^{\beta_{m+i}}_{\delta_m}
\right) \nonumber \\
&&+
\lambda_5 \,  \left( \phi^{\alpha_m}_{\beta_{m+i}}
\phi^{\gamma_{m+j}}_{\alpha_m}
\phi^{\delta_m}_{\gamma_{m+j}} \phi^{\beta_{m+i}}_{\delta_m}
+ \phi^{\alpha_m}_{\beta_{m-i}} \phi^{\gamma_{m-j}}_{\alpha_m}
\phi^{\delta_m}_{\gamma_{m-j}} \phi^{\beta_{m-i}}_{\delta_m}
\right)  \nonumber \\
&& \left.
- 2
 \lambda_2 \, \left(\phi^{\alpha_m}_{\beta_{m-j}}
\phi^{\gamma_{m+i}}_{\alpha_m}
\phi^{\delta_{m+i-j}}_{\gamma_{m+i}} \phi^{\beta_{m-j}}_{\delta_{m+i-j}}+
\phi^{\alpha_m}_{\beta_{m-i}} \phi^{\gamma_{m+j}}_{\alpha_m}
\phi^{\delta_{m-i+j}}_{\gamma_{m+j}} \phi^{\beta_{m-i}}_{\delta_{m-i+j}}
\right)\right], \nonumber \\
\eeqa
where we have used the shorthand notation $m+i=m+l_i=m+k_i+k_4$ in the above
sums. In ${\cal N}=1$ language, the $\lambda_1$, $\lambda_3$, and
$\lambda_4$ couplings are descendants of the D-terms, while the $\lambda_2$
coupling is a descendant of the superpotential term, and $\lambda_5$ receives
contributions from both terms. In our normalization $\lambda_5$ is twice
the superpotential coupling minus the D-term coupling. The Lagrangian
obtained by orbifolding the ${\cal N}=4$ theory corresponds to ``degenerate''
values of couplings: $Y^2=\lambda_1=\lambda_2=\lambda_3=\lambda_4=
\lambda_5=g^2$, where $g$ is the gauge coupling.
However, as we will see below, for these values of the couplings
 the $\beta$ functions do not
vanish. Therefore, if the theory is indeed conformal for finite $N$,
one has to assume that there will be a different set of couplings for
which all the $\beta$ functions vanish.
However, for generic values of the
quartic scalar couplings the potential is unbounded from below, while when
all couplings are identical the potential is positive definite
(as guaranteed by the supersymmetry of the theory it was projected from).
We will assume that the ratios of the couplings are sufficiently close to one
at the zeros of the $\beta$ functions so that the potential is bounded.
We will see later that this is
true in the large $N$ limit.

\section{The Renormalization Group Equations}
\setcounter{equation}{0}
\setcounter{footnote}{0}

To calculate the one-loop $\beta$ functions we rely heavily on the work of
Machacek and Vaughn~\cite{MachacekVaughn} who summarized one-loop results
and derived two-loop $\beta$ functions for a general field theory. We first
calculated the
${\cal N}=4$ SUSY $\beta$ functions for the gauge, Yukawa and quartic
couplings despite the fact that they are related by supersymmetry.
In order for this calculation to be useful for the non-supersymmetric
orbifold theories one has
to refrain from using the superfield
formalism and instead deal separately with component
scalar, fermion, and gauge boson fields. There is a term by term
correspondence between the ${\cal N}=4$ theory and the orbifolded theory in
the large $N$ limit~\cite{BershadskyJohansen}.
The fact that all the $\beta$ functions vanish when SUSY relations are
imposed between the various couplings provides strong cross checks on the
calculation.

At one-loop the gauge $\beta$ function vanishes identically \cite{VafaFramp},
so at one-loop the gauge coupling is a free parameter.  The general
one-loop $\beta$ function for the Yukawa couplings is~\cite{MachacekVaughn}:
\beqa
(4 \pi)^2 \beta^a_Y &=& {{1}\over{2}} \left[ Y^\dagger_2(F) Y^a + Y^a Y_2(F)
\right]
+ 2 Y^b Y^{\dagger a} Y^b  \nonumber \\ &&+ Y^b \, {\rm Tr} \, Y^{\dagger b}
Y^a - 3 g_m^2 \{C^m_2(F), Y^a \}
\label{betayuk}
\eeqa
where $Y^a_{ij}$ is the Yukawa coupling of scalar $a$ to fermions $i$ and $j$,
\beq
Y_2(F) = Y^{\dagger a} Y^a ~,
\eeq
and $C^m_2(F)$ is the quadratic Casimir of the fermion fields transforming
under the $m$-th gauge group. Thus the first
term in Eq.~(\ref{betayuk}) represents scalar loop corrections to the fermion
legs, the second term
1PI corrections from the Yukawa interactions themselves, the third term 
fermion loop corrections to the scalar leg, and the last term represents
gauge loop corrections to the fermion legs.

The
Yukawa $\beta$ function can be derived by projecting the ${\cal N}=4$ result
graph by graph (see the Appendix). The only changes are that $ |\Gamma| N$ is
replaced by  $N$ and the fermions are in bifundamental representations
rather than the adjoint. Thus we
find:
\beq  \label{betaYukawa}
(4 \pi)^2 \beta_Y =  6 N Y^3 - 6 \, {{N^2-1}\over{ N}} g^2 Y  ~,
\eeq
so $\beta_Y$ vanishes when
\beq
Y= Y_* \equiv g \, \sqrt{ 1- {{1}\over{N^2}} }.
\eeq
Note that this result is independent of the values of the quartic scalar
couplings.

In the notation of Machacek and Vaughn \cite{MachacekVaughn} the $\beta$
function for a quartic scalar coupling at one-loop is given by
\beq
\label{eq:quarticbeta}
(4 \pi)^2 \beta_{\lambda} = \Lambda^2 - 4H +3 A +\Lambda^Y -3\Lambda^S,
\eeq
where $\Lambda^2$ corresponds to the 1PI contribution from the quartic
interactions themselves and should not be confused with a mass scale,
$H$ corresponds to the fermion box graphs,
$A$ to the two gauge boson exchange graphs, $\Lambda^Y$ to the Yukawa
leg corrections, and finally $\Lambda^S$ corresponds to the gauge leg
corrections.  The contributions to $\Lambda^2$,  $H$,
and $\Lambda^Y$ can be found by simply projecting the ${\cal N}=4$ results
(see the Appendix).  The contributions to $\Lambda^S$ can be found by noting
that the scalars are bifundamentals rather than adjoints.
The gauge 
boson exchange term, $A$, can be calculated by a simple manipulation of
the gauge generators, which is  explained in the Appendix.
We find:
\beqa
(4 \pi)^2 \beta_{\lambda_1}  &=&N(
        4 \lambda_1^2 +
          \lambda_3^2 +2 \lambda_4^2 +2  \lambda_5^2   -16 Y^4 +16
\lambda_1 Y^2) \nonumber \\ &&+ 3{{N^2-4}\over{N}} g^4 -12{{N^2-1}\over{ N}}
g^2 \lambda_1\, , \label{betalambda1}
\\
(4 \pi)^2 \beta_{\lambda_2} &=&N(-2 \lambda_2 \lambda_4 -2 \lambda_2
\lambda_5 +8 Y^4 -16 \lambda_2
            Y^2)  +12{{N^2-1}\over{ N}} g^2 \lambda_2 \, ,
\nonumber \\ \\
(4 \pi)^2 \beta_{\lambda_3} &=&N({{1}\over{2}} \lambda_3^2 -
          2 \lambda_1 \lambda_3 +2 \lambda_4 \lambda_5    -8   \lambda_3 Y^2)
   \nonumber \\
&&+ 3{{N^2-4}\over{2 N}} g^4 +6{{N^2-1}\over{ N}} g^2 \lambda_3\, ,
\\
(4 \pi)^2 \beta_{\lambda_4} &=&N( {{1}\over{2}}
        \lambda_5^2  +2 \lambda_2^2 +\lambda_4^2  +2 \lambda_1 \lambda_4-
          \lambda_3 \lambda_5 -8 Y^4 +8 \lambda_4 Y^2)  \nonumber \\
&&+3{{N^2-4}\over{2 N}} g^4 -6{{N^2-1}\over{ N}} g^2 \lambda_4\, ,
\\
(4 \pi)^2 \beta_{\lambda_5} &=&N( {{1}\over{2}}
        \lambda_5^2    +2 \lambda_2^2 -\lambda_3 \lambda_4 +
          \lambda_4 \lambda_5 +2 \lambda_1 \lambda_5-8 Y^4 +8 \lambda_5 Y^2)
\nonumber \\
&&+3{{N^2-4}\over{2 N}} g^4 -6{{N^2-1}\over{ N}} g^2 \lambda_5\, .
\label{betalambda5}
\eeqa

Finding the general solution for the simultaneous zeroes of the
$\beta_{\lambda}$ functions is obviously a complicated problem,
here we choose to focus on
the solutions for the couplings that reduce in the large $N$ limit to the
${\cal N}=4$ SUSY fixed point, i.e. $\lambda_{i*} \rightarrow g^2$.
At order $1/N^4$ there are two such solutions which are given by:

\beqa
\lambda_{1*} &\approx& g^2 \left(1- {{5}\over{8 N^2}}  +{{459}\over{1024 N^4}}
 + \ldots \right) \nonumber \\
\lambda_{2*} &\approx& g^2 \left(1- {{19}\over{16 N^2}}  -{{387}\over{2048 N^4
}} + \ldots \right) \nonumber \\
\lambda_{3*} &\approx& g^2 \left(1- {{7}\over{4 N^2}}  -{{423}\over{512 N^4}}
 + \ldots \right) \\
\lambda_{4*} &\approx& g^2 \left(1- {{5}\over{8 N^2}}  +{{459}\over{1024 N^4}}
 + \ldots \right) \nonumber \\
\lambda_{5*} &\approx& g^2 \left(1- {{5}\over{8 N^2}}  +{{459}\over{1024 N^4}
} + \ldots \right) \nonumber
\eeqa
and
\beqa
\lambda_{1*} &\approx& g^2 \left(1- {{19}\over{16 N^2}}
+{{225}\over{8192 N^4}} + \ldots \right) \nonumber \\
\lambda_{2*} &\approx& g^2 \left(1- {{47}\over{32 N^2}}
-{{1467}\over{16384 N^4}} + \ldots \right) \nonumber \\
\lambda_{3*} &\approx& g^2 \left(1- {{5}\over{8 N^2}}
-{{153}\over{4096 N^4}} + \ldots \right) \\
\lambda_{4*} &\approx& g^2 \left(1- {{1}\over{16 N^2}}
+{{5067}\over{8192 N^4}} + \ldots \right) \nonumber \\
\lambda_{5*} &\approx& g^2 \left(1- {{1}\over{16 N^2}}
+{{5067}\over{8192 N^4}} + \ldots \right) \nonumber
\eeqa

We should note that the zeroes of the $\beta$ functions are not true fixed
points.  This is because
we have not included all possible quartic couplings allowed by gauge
invariance, we have
only included the quartic couplings that arise from the projection from
the ${\cal N}=4$ theory.  Examples of operators that do not appear in the
tree-level
Lagrangian of these orbifold theories include
\beq
\phi^{\alpha_m}_{\beta_{m+i}}  \phi^{\beta_{m+i}}_{\alpha_m}
\phi^{\gamma_m}_{\delta_{m+i}} \phi^{\delta_{m+i}}_{\gamma_m}
\ \ {\rm and}\ \
\phi^{\alpha_m}_{\beta_{m+i}}  \phi^{\beta_{m+i}}_{\alpha_m}
\phi^{\gamma_m}_{\delta_{m-i}} \phi^{\delta_{m-i}}_{\gamma_m}.
\label{extra}
\eeq
Such gauge invariant operators are induced, for example, by two gauge
boson exchange diagrams.  In the non-supersymmetric theory there is no
symmetry or non-renormalization theorem that prevents these operators
from appearing via RG evolution.  A full calculation would require
considering all possible quartic interactions,
and finding the simultaneous zeroes of all $\beta$ functions.
However, if the fixed point values of some of these new couplings are
non-zero then, as we will see, we loose the special large $N$ behavior
of the pure projected theory.

We will proceed as follows:  we assume that the effective ``conformal" theory
is embedded in a more fundamental theory at a scale $M$ (e.g.  some set of
particles of mass $M$ are integrated out of the theory at this scale),
we assume that  the theory has been arranged such that
the  $\beta$ functions for $Y$ and $\lambda_i$ vanish, and
that at this particular renormalization scale, $M$, all other quartic
couplings vanish. We can then compute the proper
1PI contribution to the mass of any particular scalar. We will
only keep the quadratically divergent piece.

The quadratic divergence is given by
\beq
m^2_\phi = \left[N(2 \lambda_1  - \lambda_3 + 2 \lambda_4 + 2 \lambda_5 ) +
3{{N^2-1}\over{N}} g^2 -8 Y^2 N \right] \int^M{{d^4p}\over{(2 \pi)^4}}
{{1}\over{p^2}}.
\eeq
Plugging in our solutions for the zeroes of the $\beta$ functions we have (to
lowest non-vanishing order in $N$) for both cases:
\beq
\label{eq:quad}
m^2_\phi=  {{3 g^2}\over{N}}  {{M^2}\over{16 \pi^2}}.
\eeq
Note that, as expected, the terms linear in $N$ canceled.
Thus we see that there is a technically unnatural hierarchy in this set of
theories.
In order to keep the scalars light a mass counterterm must be tuned, order by
order,
to cancel quadratic divergences. Alternatively, $N$ has to be taken
extremely large. For $m=m_{weak}\approx 1$ TeV, $M=M_{Pl}\approx 10^{18}$ GeV,
$\frac{g^2}{4\pi}\approx \frac{1}{30}$
we find that one would need $N \approx 10^{28}$.

We now briefly comment on the possible effects of including other quartic
operators like those displayed in Eq. (\ref{extra}).  There is a contribution
to $\Lambda^2$ of (\ref{eq:quarticbeta}) of
order $N^2 (\lambda^{{\rm new}})^2$, the contribution
to $A$ is of order $g^4$ (see Appendix).
Thus the form of the $\beta$ function is:
\beqa
(4 \pi)^2 \beta^{\rm new}_k  &=&
     N^2 a^{ij}_k \lambda^{\rm new}_i \lambda^{\rm new}_j  +
     N b^{ij}_k \lambda^{\rm new}_i  \lambda_j +
     c^{ij}_k \lambda_i \lambda_j +16 N  \lambda^{\rm new}_k  Y^2 \nonumber \\
 && +d_k 3(1 + {{2}\over{N^2}}) g^4
    -12{{N^2-1}\over{ N}} g^2 \lambda^{\rm new}_k ~,
\eeqa
where we have taken the coupling $\lambda^{\rm new}_k$ to have the same sign
and
normalization as $\lambda_1$. In the above formula,
$d_k$ is an integer, depending on how many gauge groups the scalar
fields share (see Appendix).
Thus we expect $\lambda^{\rm new}_k$ to be of
order $g^2/N$ at a fixed point.
The contribution of the graphs arising from these operators to the quadratic
divergence is of order $N^2$, so the contribution to $m^2_\phi $ is of order
$g^2 N$.  Thus the inclusion of these additional operators seems to make
the naturalness problem much worse.
It may be possible to cancel
the quadratic divergence order by order, but a priori there seems to be no
reason for such a cancellation to occur at a fixed point of the theory.

Using the methods presented above one can also calculate the
two-loop gauge $\beta$ function. The two-loop piece of the gauge
$\beta$ function in a general gauge theory is given by \cite{MachacekVaughn}:
\begin{eqnarray}
\beta^{(2)}_g=&&-\frac{g^3}{(4\pi)^4} \left[ \left\{ \frac{34}{3} (C_2(G))^2
-\frac{1}{2} \left( 4 C_2(F)+\frac{20}{3}C_2(G)\right)S_2(F) \right. \right.
\nonumber \\&&-\left. \left.
\left( 4 C_2(S)+\frac{2}{3}C_2(G)\right)S_2(S)\right\}g^2+Y_4(F)\right],
\label{2loopbeta}
\end{eqnarray}
where $C_2(G)$ is the Casimir of the adjoint, $C_2(F) S_2(F)$ is the
sum over (two-component) fermions of the Casimir times the Dynkin index
in the given representation,  $C_2(S) S_2(S)$ is the same for  complex scalars,
while $Y_4(F)$ is the contribution of the Yukawa couplings defined by
\begin{equation}
{\rm Tr} Y^a Y^{\dagger a} t^A t^B=Y_4(F) \delta^{AB},
\end{equation}
where $Y^a$ are the Yukawa coupling matrices  for the scalar field  $a$,
and $t^A$ are the gauge generators in the representation of the fermion
fields.

For the orbifold theory considered above these expressions are given by
\begin{eqnarray}
&& C_2(G)=N, \nonumber \\
&& C_2(F)S_2(F)=4 (N^2-1), \nonumber \\
&& C_2(G)S_2(F)=4N^2, \nonumber \\
&& C_2(S)S_2(S)=3(N^2-1), \nonumber \\
&& C_2(G)S_2(S)=3N^2, \nonumber \\
&& Y_4(F)=24N^2Y^2.
\end{eqnarray}
Note that Eq.~(\ref{2loopbeta}) is independent of the quartic scalar couplings.
At the one-loop fixed point of the Yukawa coupling, which is also independent
of the values of the quartic scalar couplings, $Y^2=\frac{N^2-1}{N^2}g^2$.
Using this value we find that the leading order terms in $N$ cancel,
and the sub-leading pieces give
\begin{equation}
\beta^{(2)}_g=\frac{4g^5}{(4\pi)^4} >0,
\end{equation}
thus the theory is not asymptotically free. If the theory is indeed
conformal, then the fixed point would necessarily be a UV fixed point.
In order to check whether the theory is conformal or not, one would need
to study the three-loop gauge $\beta$ function. If the three-loop term
turns out to be negative and of ${\cal O}(N^2)$, then there will be a
perturbative UV fixed point, since the fixed point will be
$g^2 ={\cal O}(1/N^2)$ and higher loop corrections to the gauge $\beta$
function can be neglected. For any other case there cannot
be a perturbative fixed point. For example if the three-loop term is
${\cal O}(N)$, then any putative fixed point can only be seen by summing
all planar diagrams. Such a fixed point could exist independent of the sign
of the three-loop term.

If this theory turns out to be conformal with a perturbative fixed point,
then this could provide an interesting
example of a theory with a non-trivial UV
fixed point. Such a theory could then serve as a counter example
to the conjecture presented in Ref.~\cite{ACS}.

\section{Conclusions}
\setcounter{equation}{0}
\setcounter{footnote}{0}

In this paper we have considered a particular class of non-supersymmetric
orbifold theories obtained from finite ${\cal N}=4$ theories.
Our calculations are summarized by equations (\ref{betaYukawa}),
(\ref{betalambda1})--(\ref{betalambda5}) and (\ref{eq:quad}). 
We calculated the
one-loop $\beta$ functions
and found the simultaneous zeroes that approach the SUSY fixed point
in the large $N$ limit. At  one-loop the theory possesses quadratic divergences
in sub-leading orders in $N$ and therefore cannot stabilize the
weak scale without $N$ being unreasonably large.

RG running also generates new operators (quartic scalar
couplings) which are not present in the tree-level orbifold Lagrangian.
These new couplings will shift the fixed point values of the original
operators,
and also contribute to the quadratic divergences themselves.
It is possible, but unlikely, that with these new couplings all
quadratic divergences vanish. The difficulty in canceling the
quadratic divergences stems from the fact that the contributions of
the new operators to the quadratic divergence is more important in the
$1/N$ expansion than the divergences we have discussed here. We think
that a cancellation is unlikely to occur, but the importance of the
problem merits further investigation which would involve the renormalization
of the full set of operators allowed by symmetries.

\section*{Acknowledgments}
We are grateful to Nima Ar\-ka\-ni-Ha\-med,
Bob Cahn, Andre de Go\-u\-ve\-a, Mi\-cha\-el Graesser, Shamit Kachru, 
Chris Kolda,
Aneesh Manohar, Hitoshi Murayama, Markus Luty, Lisa Randall,
Martin Schmaltz, and Raman Sundrum
for useful discussions. W.S. thanks the members of the theory group at
Berkeley for their  hospitality during a visit when this work was initiated.
C.C. and J.T. are
supported in part by the U.S. Department of Energy under Contract
DE-AC03-76SF00098 and in part by the National Science Foundation under
grant PHY-95-14797. W.S. is supported in part by the U.S. DOE under Contract
DOE-FG03-97ER40506.
C.C. is a research fellow of the Miller Institute
for Basic Research in Science.

\appendix
\renewcommand{\thesection}{Appendix \Alph{section}}
\section{${\cal N}=4 \,\, \beta$ Functions}
\renewcommand{\theequation}{\Alph{section}.\arabic{equation}}
\setcounter{equation}{0}
\setcounter{footnote}{0}

${\cal N}=4$ supersymmetric theories are finite, therefore the $\beta$
function vanishes to all orders in perturbation theory. In terms of component
fields the ${\cal N}=4$ Lagrangian has three different kinds of couplings:
gauge,  Yukawa, and  quartic scalar. Even though
these couplings are related by ${\cal N}=4$ supersymmetry it is
useful to calculate their $\beta$ functions separately. In the orbifold
theories different couplings are not related by supersymmetry, yet
${\cal N}=4$ results are helpful in the calculation of the non-supersymmetric
$\beta$ functions.

The ${\cal N}=4$ theory can be thought of as an ${\cal N}=1$ theory with three
adjoint chiral superfields and a superpotential for these fields.
When the ${\cal N}=4$ theory is expressed in terms of ${\cal N}=1$ component
fields the $SU(4)_R$ global symmetry is not explicit in the Lagrangian,
only its $SU(3)\times U(1)$ subgroup is manifest.
In terms of components the Lagrangian is given by
\begin{eqnarray}
\label{eq:N=4}
&&{\cal L}_{N=4}=
-\frac{1}{4}F_{\mu\nu}F^{\mu\nu}-i \bar{\lambda}^a \sigma^{\mu}
D_{\mu} \lambda^a -i \bar{\Psi}_i^a \sigma^{\mu} D_{\mu} \Psi_i^a
+D^{\mu}\phi^{\dagger  a}_i D_{\mu}\phi_i^a+ \nonumber \\
&& -\sqrt{2}g f^{abc}(\phi^{\dagger  c}_i \lambda^a \Psi^b_i -
\bar{\Psi}^c_i  \bar{\lambda}^a \phi^b_i)
-\frac{Y}{\sqrt{2}}\epsilon_{ijk} f^{abc}
(\phi^{c}_i \Psi_j^a \Psi^b_k +
\bar{\Psi}^c_i \bar{\Psi}^a_j \phi^{\dagger \,b}_k)\nonumber \\
&& +\frac{g^2}{2} (f^{abc}\phi^b_i\phi^{\dagger \, c}_i)
(f^{ade}\phi^d_j\phi^{\dagger \, e}_j)-\frac{Y^2}{2}
\epsilon_{ijk}\epsilon_{ilm} (f^{abc} \phi^b_j\phi^c_k)(f^{ade}
\phi^{\dagger d}_l  \phi^{\dagger e}_m),
\end{eqnarray}
where $a, \ldots ,e =1,\ldots,N^2-1$ are the adjoint gauge indices,
while $i, \ldots, m=1,2,3$ are $SU(3)$ flavor indices.
The $SU(N)$ structure constant is denoted by $f^{abc}$, $\lambda$ is the
(two-component) gaugino, $\Psi_i$ are the (two-component) adjoint
fermions, and $\phi_i$ are the three complex adjoint  scalars.
Meanwhile $g$ is the gauge coupling and $Y$ is the coupling of
the superpotential term  for the chiral superfields.
The above Lagrangian is ${\cal N}=4$ supersymmetric
for $Y=g$. In order to easily identify the origin of different terms
in the calculation it is instructive to keep $Y$ explicit
in the Lagrangian.

\begin{figure}
\PSbox{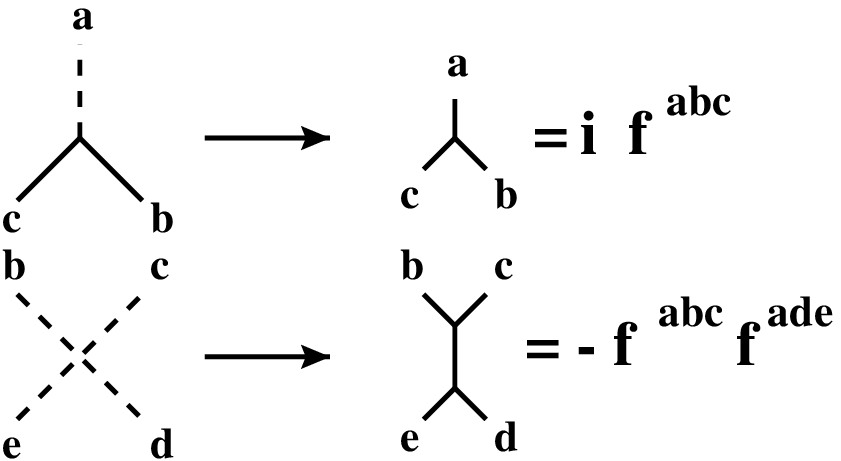 hscale=75 vscale=75 hoffset=100  voffset=0}{13.7cm}{4cm}
\caption{The group theory Feynman diagrams for the Yukawa couplings and the
quartic scalar couplings of the ${\cal N}=4$ theory.\label{fig:rules}}
\end{figure}

The one-loop (as well as two-loop) $\beta$ functions are known
for a  general field theory \cite{MachacekVaughn}. In order to use the
formulae given in Ref.~\cite{MachacekVaughn} one needs to calculate
certain group theoretic factors.
This calculation can be conveniently carried out using the method of
Cvitanovic~\cite{Cvitanovic},
in which one draws a separate ``group theory diagram'' for every
Feynman diagram. Evaluating these group theory diagrams will then amount to
calculating the group theory coefficients needed for the general formulas
of the $\beta$ functions of \cite{MachacekVaughn}. Since all fields are in
the adjoint representation every Yukawa coupling carries a
factor $f^{abc}$ while every quartic scalar coupling carries a factor
$f^{abc} f^{ade}$. In order to obtain the group theory diagrams
one replaces every factor of $i f^{abc}$ with a cubic vertex
(see Fig.~\ref{fig:rules}).
The diagram obtained this way does not have to coincide with the actual
form of the Feynman diagram that one is evaluating.

Using the Lagrangian (\ref{eq:N=4}) and the above rules of calculating
the group theory factors one can obtain the various $\beta$ functions for
the ${\cal N}=4$ theory. The one-loop $\beta$ function for the gauge coupling
is  given by
\begin{equation}
(4\pi)^2 \beta_{g}= -g^3(\frac{11}{3}C_2 (G)-\frac{2}{3} S_2 (F)
-\frac{1}{3} S_2 (S)),
\end{equation}
where $C_2(G)$ is the Casimir of the adjoint, $S_2(F)$ is the Dynkin index
of the (two-component) fermions, and $S_2(S)$ is the Dynkin index for the
complex scalars. For the ${\cal N}=4$ theory $C_2(G)=N, S_2(F)=4 N, S_2(S)=3N$,
and thus $\beta_{g}=0$.

The one-loop $\beta$ function for the Yukawa coupling $Y^a$ in a general gauge
theory is given by the formula
\begin{eqnarray}
(4 \pi )^2 \beta_{Y}^a=&& \hspace*{-0.5cm}\frac{1}{2} 
(Y_2^{\dagger}(F)Y^a+Y^a Y_2(F))+
Y^bY^{\dagger a}Y^b+ Y^b {\rm Tr} Y^{\dagger b} Y^a
\nonumber \\ &&\hspace*{-0.5cm} -3g^2 \{ C_2(F),Y^a \}.
\end{eqnarray}
In the case of the ${\cal N}=4$ theory we evaluate the $\beta$ function of the
vertex $-\sqrt{2}g f^{abc}\phi^{\dagger  c}_i \lambda^a \Psi^b_i$. In the
projected orbifold theory all Yukawa couplings are equal due to the
$Z_k$ symmetry of the theory, thus we can use any of the ${\cal N}=4$
vertices to obtain the projected result. For this coupling the different
terms in the  above $\beta$ function are:
\begin{eqnarray}
&& \frac{1}{2} (Y_2^{\dagger}(F)Y^a+Y^a Y_2(F))=(4Ng^2+2NY^2) \sqrt{2}g,
\nonumber \\
&&  Y^bY^{\dagger a}Y^b= (-4NY^2)\sqrt{2}g, \nonumber \\
&& Y^b {\rm Tr} Y^{\dagger b} Y^a=(2Ng^2+2NY^2)\sqrt{2}g, \nonumber \\
&& -3g^2 \{ C_2(F),Y^a \} =(-6Ng^2)\sqrt{2}g.
\end{eqnarray}
The sum of these terms adds up to zero independently of the value of
$Y$, which can be understood in the following way: for $Y\neq g$ we have
an ${\cal N}=1$ supersymmetric theory with three adjoint fermions and a
non-vanishing superpotential. Since we have chosen the $\beta$ function of the
Yukawa coupling involving the gaugino, therefore the Yukawa $\beta$
function has to be proportional to the gauge $\beta$ function for
any value of $Y$. The one-loop $\beta$ function of the gauge coupling is
independent of $Y$ therefore the cancellation has to happen for a generic
value of $Y$. This provides an independent check of our result.

\begin{figure}
\PSbox{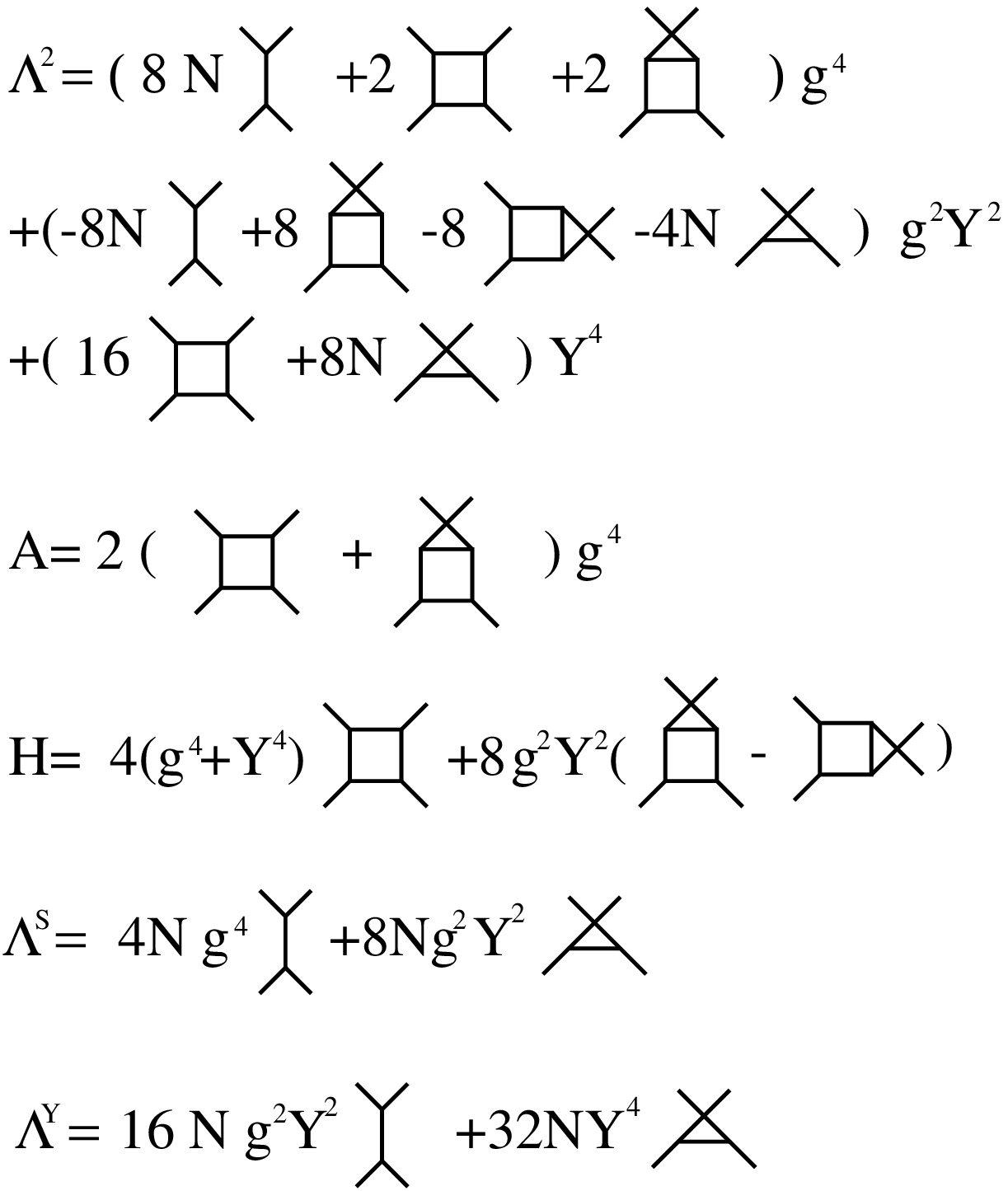 hscale=50 vscale=50 hoffset=110  voffset=0}{13.7cm}{6cm}
\caption{Contributions to the $\beta$ function of the quartic scalar
couplings of the fields
$\phi_1^a \phi^{\dagger b}_1  \phi_2^c \phi^{\dagger d}_2$ in the ${\cal N}=4$
theory. The ordering of the fields in the above diagrams is clockwise,
with $\phi_1^a$ in the upper left corner. The meaning
of the above group theory diagrams is explained in
Fig.~\protect\ref{fig:rules}.
\label{fig:beta1}}
\end{figure}

\begin{figure}
\PSbox{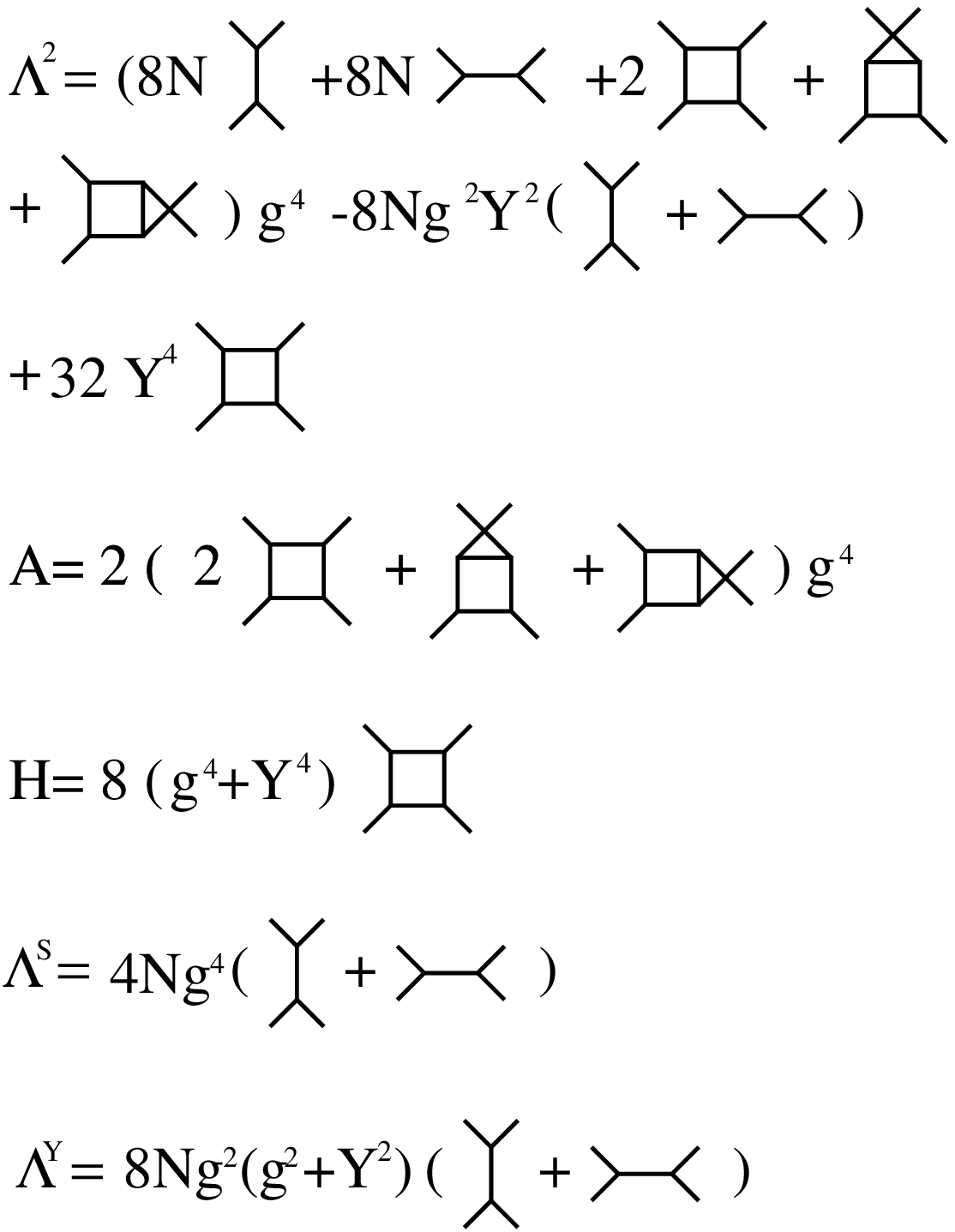 hscale=50 vscale=50 hoffset=110  voffset=0}{13.7cm}{7.5cm}
\caption{Contributions to the $\beta$ function of the quartic scalar
couplings of the fields
$\phi_1^a \phi^{\dagger b}_1  \phi_1^c \phi^{\dagger d}_1$ in the ${\cal N}=4$
theory.  The ordering of the fields in the above diagrams is clockwise,
with $\phi_1^a$ in the upper left corner. The meaning
of the above group theory diagrams is explained in
Fig.~\protect\ref{fig:rules}.
\label{fig:beta2}}
\end{figure}

\begin{figure}
\PSbox{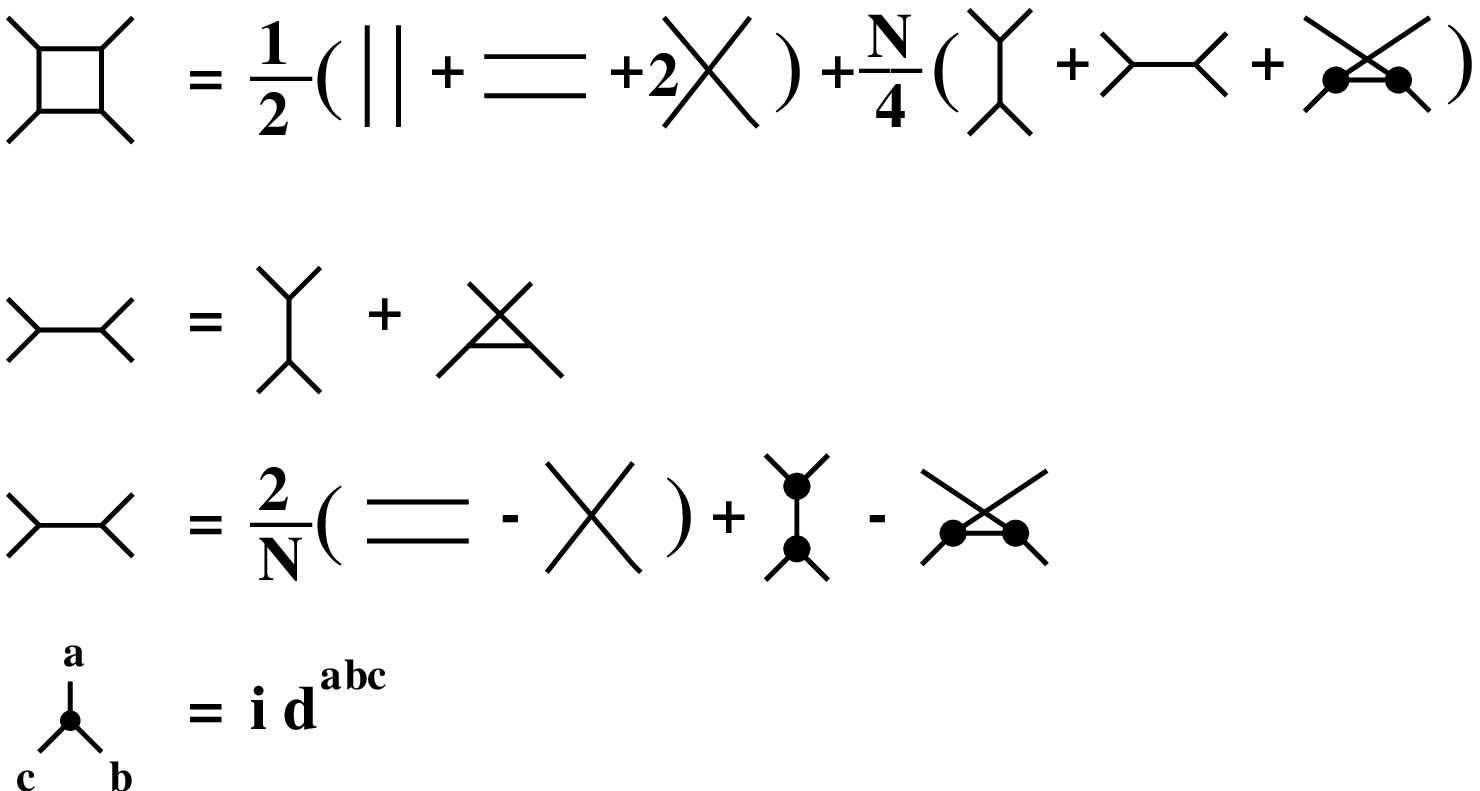 hscale=75 vscale=75 hoffset=50  voffset=0}{13.7cm}{7.5cm}
\caption{The diagrammatic representation of the $SU(N)$ group theory
identities needed to show that the ${\cal N}=4$ $\beta$ functions
of the quartic couplings do indeed vanish. The first line gives
the decomposition of the ``gluon box diagram'' in terms of a complete
set of tensors, the second line is the Jacobi identity, while
the third line is an identity relating different combinations of the
$d$ and $f$ tensors. A single unconnected line corresponds to $\delta^a_b$.
These results are taken from \protect\cite{Cvitanovic}.
\label{fig:box}}
\end{figure}

Finally we calculate the one loop $\beta$ functions for the quartic scalar
couplings. The general formula for an arbitrary gauge theory is given by
\begin{equation}
\label{eq:quartic}
(4 \pi )^2 \beta_{\rm quartic}= \Lambda^2+3A-4H+\Lambda^Y-3\Lambda^S.
\end{equation}
We calculate two different combinations of quartic $\beta$ functions in
the ${\cal N}=4$ theory: one for the coupling of the operator
$\phi_1^a \phi^{\dagger b}_1\phi_2^c \phi^{\dagger d}_2$, for which the
contributions are given in Fig.~\ref{fig:beta1}, and another for the
operator $\phi_1^a \phi^{\dagger b}_1  \phi_1^c \phi^{\dagger d}_1$
the contributions to which are given in Fig.~\ref{fig:beta2}.
Combining these results according to Eq.~(\ref{eq:quartic})
one finds that these $\beta$ functions indeed vanish
for the ${\cal N}=4$ theory.  Cancellation of various terms occurs after
decomposing the ``gluon box'' diagrams \cite{Cvitanovic}
in a complete basis of group theory tensors using the results
given in Fig.~\ref{fig:box}.

\begin{figure}
\PSbox{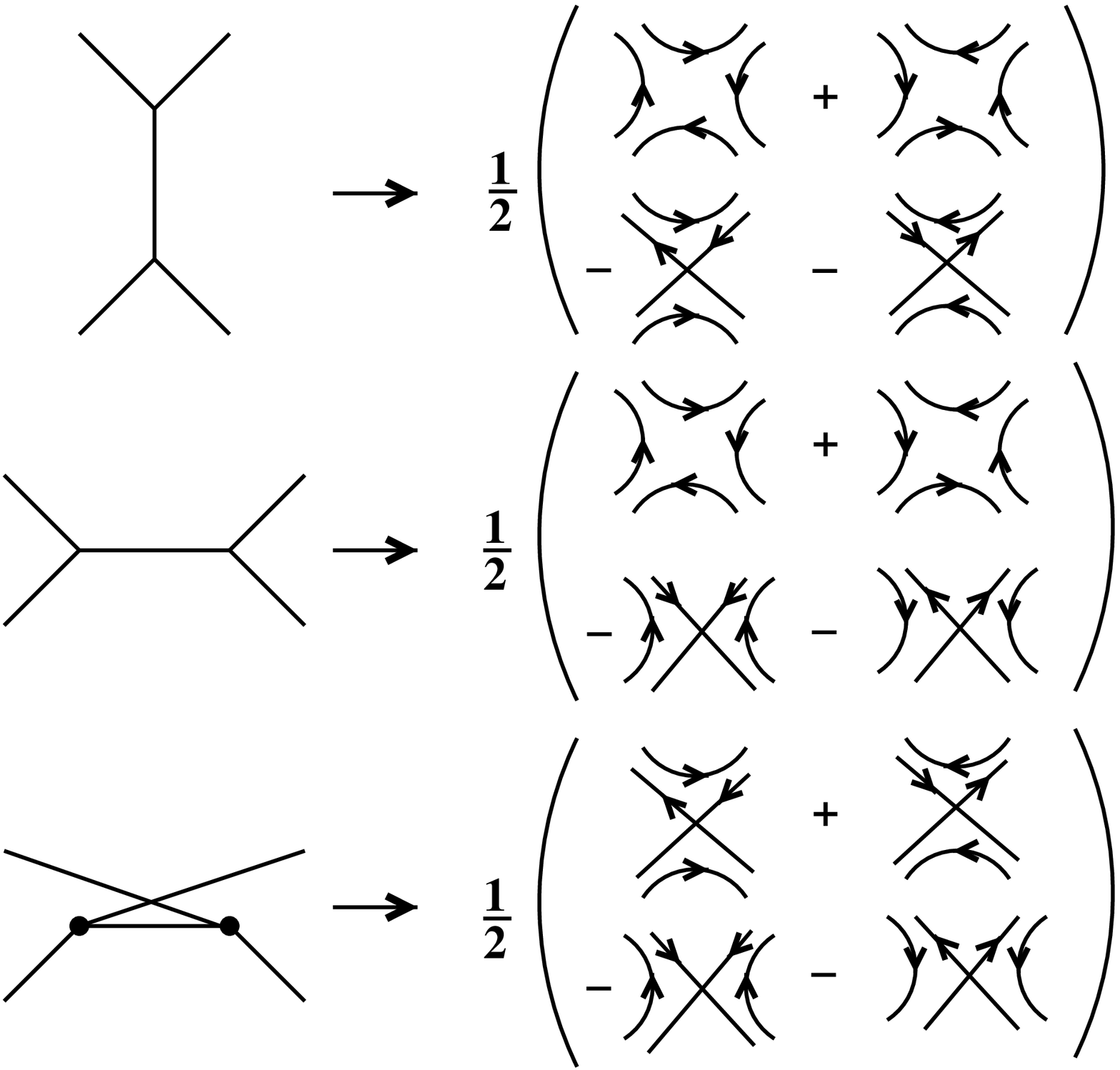 hscale=50 vscale=50 hoffset=60  voffset=0}{13.7cm}{7cm}
\caption{The large $N$ rules for adjoints.
\label{fig:largeN}}
\end{figure}

In order to project the ${\cal N}=4$ theory down to the orbifolded theory it
is convenient to make use of large $N$ double-line notation, since all our
fields are bifundamentals. To do this we need two $SU(N)$ identities:
\beq
i f^{abc}= 2 {\rm Tr} \, (T^a T^b T^c -T^c T^b T^a) ~,
\eeq
and
\beq
(T^a)^i_j (T^b)^m_n = {{1}\over{2}} (\delta^i_n \delta^m_j -{{1}\over{N}}
\delta^i_j \delta^m_n) ~.
\label{gluonprop}
\eeq
To keep the fields canonically normalized after changing from the single
index basis to  the double index basis we need to rescale
\beq
\phi^a = \sqrt{2} \phi^j_i (T^a)^i_j ~.
\eeq
Using these identities and representing $\delta^i_j$ by a line with an arrow
we can obtain
the large $N$ results given in Fig. \ref{fig:largeN}.

\begin{figure}
\PSbox{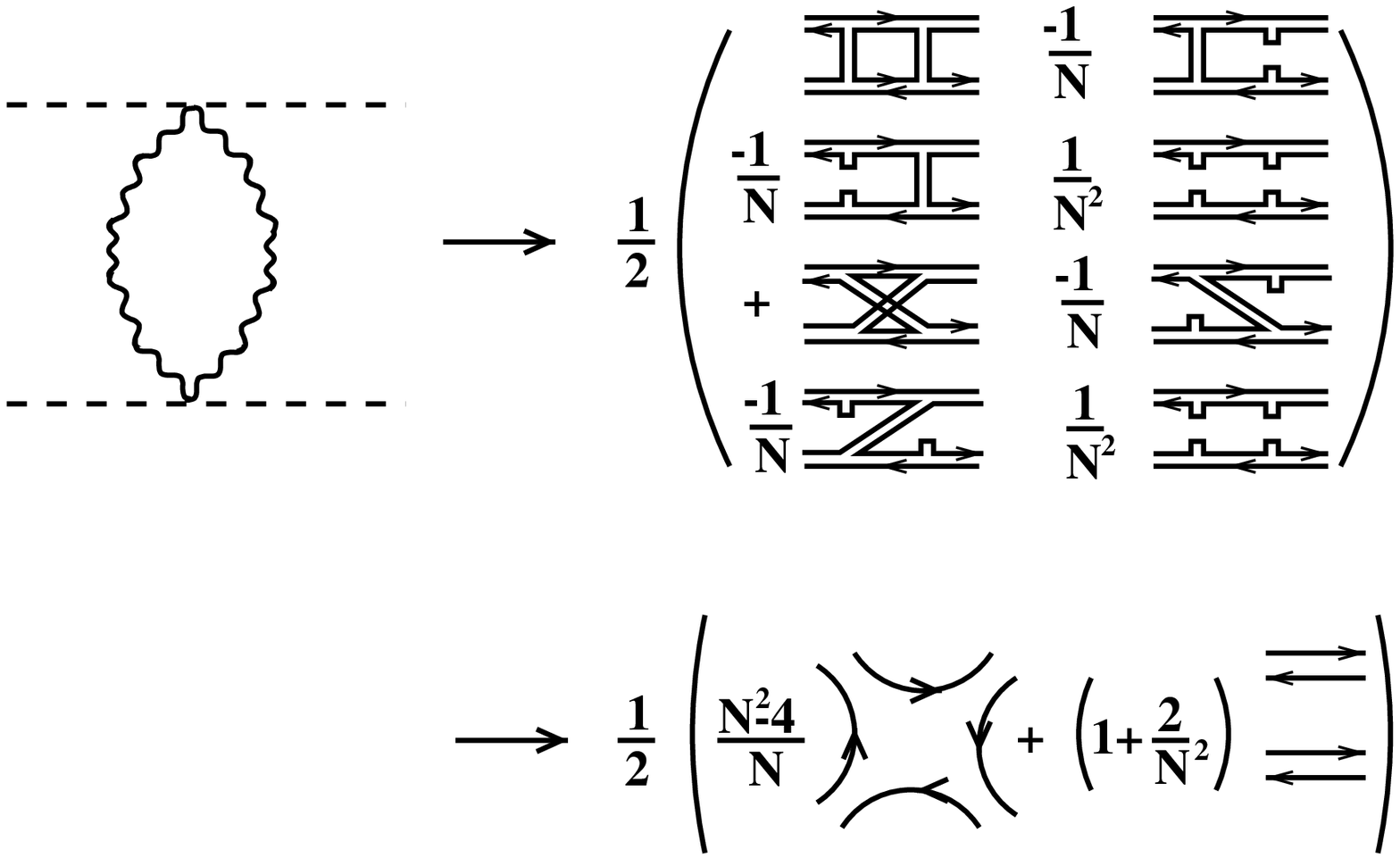 hscale=50 vscale=50 hoffset=30  voffset=0}{13.7cm}{6cm}
\caption{The proper correction to quartic couplings from gauge boson exchange.
\label{fig:gauge}}
\end{figure}

At tree-level the effect of orbifolding is similar to taking the
above large $N$ limit, the only difference is that different oriented lines
can correspond to different gauge groups.
The appropriate combination of gauge groups for each vertex can be read off
from the projected Lagrangian (\ref{Lagrangian}).  Once we have the tree-level
vertices we can simply calculate all the diagrams relevant to the $\beta$
functions.  Additionally we can apply
the projection rules to the ${\cal N}=4$ diagrams involving quartic or Yukawa
couplings, however sub-leading terms in $N$ can be generated in loops, and
these terms must be kept.  This procedure provides a check on the
calculation.

The double line notation is also convenient for gauge diagrams, however $1/N$
terms are already present in the gauge boson propagator so a little more care
must be taken. We illustrate the use of the double line notation
in the calculation of the proper correction to the quartic coupling from
two gauge boson exchange.  For simplicity, we consider the case of two
different scalar fields that share one gauge group. The calculation
proceeds by using the identity (\ref{gluonprop}) and is depicted in Fig.
(\ref{fig:gauge}).

\end{document}